\documentclass[12pt]{article}
\usepackage{amsfonts}
\usepackage{amsmath}
\catcode`@=11
\@addtoreset{equation}{section}
\catcode`@=12

\newtheorem{theo}{Theorem}

\newtheorem{lem}{Lemma}[section]

\newtheorem{prop}{Proposition}[section]

\newcommand\eps\varepsilon
\newcommand\ph\varphi
\newcommand\kap\varkappa

\renewcommand{\Re}{\mbox{\rm Re}\,}
\renewcommand{\Im}{\mbox{\rm Im}\,}

\begin{document}
\title{On analyticity of incompressible viscid fluid flow in n-dimensional torus}
\author{ O. Zubelevich \thanks{This research was partially supported by
 grants: INTAS 00-221,  RFFI  02-01-00400.}
\\ {\em Department of Differential Equations,}
\\ {\em Moscow State Aviation Institute,}
\\ {\em Volokolamskoe Shosse 4, 125871, Moscow, Russia.}
\\ {email: ozubel@yandex.ru}}

\date{ \ \ \ }
 \maketitle
\begin{abstract}We prove that the evolutionary Navier-Stokes
equation in n-D torus with initial data in the class of distributions has
an unique solution (local in t)  that is analytic by all variables. This
solution presents as a series globally.
\end{abstract}

\section{Introduction}
The regularity problems for NSE with the condition of incompressibility
have been studied by many authors. For instance, J. Serrin \cite{Serrin}
showed that under rather moderate assumptions, weak solutions are
$C^\infty$ in the space variables in the case of a conservative external
force. On the other hand, it was shown (Masuda) \cite{Masuda}, \cite{Masuda2} that if
external force is analytic in spatial variables and $t$, then srong
solutions satisfying the zero Dirichlet boundary condition are also
analytic  in the spatial variables and $t$. C. Kahane \cite{Kahane} showed
that solutions of NSE (with the incompressibility assumption and without any
boundary condition)
is analytic in spatial variable in the case of a conservative external
force.

Note that if external force and initial data are
analytic in all variables the existence of analytic solutions in all variables
can be derived from the result announced in the end of the paper
\cite{Nishida}.

In the present paper we establish the existence of analytic (in all variables) solutions under
very irregular initial conditions and obtain a global expansion of the solution
into a series.

The problem is studied in the case of periodical boundary conditions.

The author wishes to thank A. L. Skubachevski\v{\i} and
V. A. Solonnikov for useful discussions.

\section{Main theorem}
Consider the Navier-Stokes equation with initial data and the condition of incompressibility
of a fluid:
\begin{align}
\textbf{v}_t+(\textbf{v},\nabla) \textbf{v}&=-\nabla p+\nu \Delta
\textbf{v},\label{N-S-1}\\
\mbox{div}\,\textbf{v}&=0,\label{N-S-12}\\
\textbf{v}\mid_{t=0}&=\hat{\textbf{v}}(x),\label{N-S-11}
\end{align}
where by $\textbf{v}=(v^1(t,x),\ldots,v^n(t,x))$ we denote a vector-function,
$p(t,x)$ is a scalar function and $\nu$ is a positive constant.
Differential operators are defined in the usual way:
$\partial_j=\partial/\partial x_j$,
$$
\begin{array}{rclrcl}
\nabla p&=&(\partial_1 p,\ldots,\partial_n p),&
\mbox{div}\,\textbf{v}&=&\sum_{k=1}^{n}\partial_k v^k,\\
\Delta f&=&\mbox{div}\,\nabla f,&
(\textbf{v},\nabla )f&=&\sum_{k=1}^{n}v^k\partial_k f.
\end{array}
$$
Application of any scalar operator to a vector-function implies
that this operator applies to each component of the vector-function.

Note that the function $p$ is taken in problem (\ref{N-S-1})-(\ref{N-S-12}) such that
the substitution $p\mapsto p+c(t)$ ($c(t)$ is an arbitrary function) does not change the
equations. So we will find the function $p$ just up to an additional function
of $t$.
Let $k,x\in \mathbb{C}^n$, introduce some notations:
$$
(k,x)=k_1x_1+\ldots+k_nx_n,\quad
|x|=|x_1|+\ldots+|x_n|,\quad i^2=-1.$$
Sometimes we will use the Euclidian norm:
$|x|_e^2=|x_1|^2+\ldots+|x_n|^2$, and the norm
$|x|_m=\max_{k}|x_k|$.
As it is well known there is a constant $\underline{c}$ such that
$\underline{c}|x|\le|x|_e\le |x|$.

Let $\mathbb{T}^n=\mathbb{R}^n/(2\pi\mathbb{Z})^n$ be an $n-$dimensional
torus.
Define sets
$$
\begin{aligned}
\Upsilon_r(T)=&
\Big\{(t,x)\mid0<|\Im t|<\Re t<T,\quad 0<\frac{3}{\nu\underline{c}} |\Im x|_m<r<\Re t,\\
&\Re x\in\mathbb{T}^n\Big\},\\
\Upsilon (T)=&\bigcup_{0<r<T}\Upsilon_r(T).
\end{aligned}$$

For $s\in \mathbb{R}$ define a space:
$$\mathcal{H}_{s}=\Big\{f(x)=\sum_{k\in \mathbb{Z}^n}f_ke^{i(k,x)}\mid
\|f\|_{s}=\sum_{k\in
\mathbb{Z}^n}|f_k||k|^{s}<\infty\Big\}.$$

An element $f$ of the space $\mathcal{H}_{s}$ is a distribution which value of
a function $\varphi(x)\in C^\infty(\mathbb{T}^n)$ is
$$(f,\varphi)=\sum_{k\in
\mathbb{Z}^n}f_k\int\limits_{\mathbb{T}^n}e^{i(k,x)}\varphi(x)\,dx.$$
In case of positive integers $s$ the spaces $\mathcal{H}_{s}$ coincides
with the Sobolev spaces with the same subscripts \cite{Danford}.

We assume that the initial vector field $\hat{\textbf{v}}(x)$ belongs to
the space $\mathcal{H}_{s}$  and
$\mbox{div}\,\hat{\textbf{v}}(x)=0.$

Denote by $\mathcal{O}(D)$ the space of holomorphic functions in a domain
$D$.
Let $$U_{\underline{t},T}(\alpha)
=\{t\mid0<\underline{t}<\alpha |\Im t|<\Re t< T\}.$$

\begin{theo}
\label{main_theo}\begin{enumerate}\item\label{st1}
There is a positive constant $T$ such that
initial problem (\ref{N-S-1})-(\ref{N-S-11}) has an unique solution
$(p(t,x),\textbf{v}(t,x))$ in the
space $\mathcal{O}(\Upsilon (T))$.

The solution expands to a series:
$$\mathbf{v}(t,x)=\sum_{k\in
\mathbb{Z}^n}\mathbf{v}_k(t)e^{i(k,x)-t|k|_e\nu/2},$$
where  the functions $\mathbf{v}_k$ belongs to the set
$\mathcal{O}(U_{\underline{t},T}(\alpha))$ for any $\underline{t}<T$ and
$\alpha>1$,
 the
series
$$\sum_{k\in
\mathbb{Z}^n}|\mathbf{v}_k(t)||k|^s$$
uniformly converges in $\overline{U_{\underline{t},T}(\alpha)}$ for any $\underline{t}<T$ and
$\alpha>1$.

The following equalities hold:
\begin{equation}
\label{poiuyt}
\begin{aligned}
\lim_{t\to 0} \Big\|\frac{\partial^{|k|}}{\partial^{k_1}x_1\ldots\partial^{k_n}x_n}\textbf{v}(t,x)-
\frac{\partial^{|k|}}{\partial^{k_1}x_1\ldots\partial^{k_n}x_n}\hat{\textbf{v}}(x)\Big\|_{s-|k|}&=0,\\
\lim_{t\to 0} \Big\|\frac{\partial^{|k|}}{\partial^{k_1}x_1\ldots\partial^{k_n}x_n}p(t,x)-
\frac{\partial^{|k|}}{\partial^{k_1}x_1\ldots\partial^{k_n}x_n}\hat p(x)\Big\|_{s-|k|}&=0,
\end{aligned}\end{equation}
where $\hat p(x)\in\mathcal{H}_{s}$ and the vector $k$ consists of non-negative integers.
\item\label{st2} There exists positive  constant $\mu$ such that if
 $\|\hat{\mathbf{v}}\|_s\le\mu$ then the first part of this theorem remains valid
for $T=\infty$.
\end{enumerate}
\end{theo}
Note that $\mathbf{v}_0$ actually does not depend on $t$ \cite{Frisch}.
As a simple corollary of the second part of the theorem we obtain
\begin{prop}
For any $t\in\overline{U_{\underline{t},\infty}(\alpha)}$
we have
$$\max_{x\in\mathbb{T}^n}|\mathbf{v}(t,x)-\mathbf{v}_0|\le ce^{-\frac{\nu\mathrm{Re}\,t}{2}},$$
positive constant $c$ depends only on $\underline{t}$ and $\alpha$.
\end{prop}

Before approaching to a proof of the theorem we must develop
some technique tools.

\section{Definitions and technique tools}
We will denote inessential
constants by $c,C$ or by these letters with subscripts.

Provide the space $\mathcal{O}(D)$ by a collection of norms: let $u\in
\mathcal{O}(D)$ and $K$ is a compact subset of $D$ then
$\|u\|_K=\sup_{z\in K}|u(z)|$. These norms make the space $\mathcal{O}(D)$
a seminormed space. A sequence $u_k\in \mathcal{O}(D)$ converges to $u\in
\mathcal{O}(D)$ if for any compact  $K$ we have $\|u_k-u\|_K\to 0$ when $k\to
\infty$. This kind of convergence is referred to as compact convergence.

The compact convergence in the space $\mathcal{O}(\Upsilon (T))$
follows from the convergence with respect to the norms
$\|\cdot\|_{\Upsilon_r(T)}$. Indeed, every compact subset of
$\Upsilon (T)$ can be covered by finite collection
$\{\Upsilon_{r_i}(T)\}_{i=1\ldots m}$.

We say that a subset $M$ of $\mathcal{O}(D)$ is bounded if for any compact
$K$ there is a constant
$C_K$ such that for any $u$ of $M$ we have $\|u\|_K\le C_K$.

\begin{theo}[Montel]
\label{Montel}
If $M$ is a closed and bounded subset of $\mathcal{O}(D)$ then it is a
compact.
\end{theo}

Denote by $\mathcal{ \tilde O}(\Upsilon (T))$ a subset of
$\mathcal{O}(\Upsilon (T))$
that consists of functions $u$ with zero mean value:
$\int_{\mathbb{T}^n}u(t,x)\,dx=0$.

The Laplace operator $\Delta:\mathcal{\tilde O}(\Upsilon (T))\to
\mathcal{\tilde O}(\Upsilon (T))$ is invertible: there exists a bounded operator
$\Delta^{-1}:\mathcal{\tilde O}(\Upsilon (T))\to
\mathcal{\tilde O}(\Upsilon (T))$.
If $$
u(t,x)=\sum_{k\in\mathbb{Z}^n\backslash\{0\}}u_k(t)e^{i(k,x)}$$ is the
Fourier expansion of $u$ then an explicit
form of this operator is
\begin{equation}
\label{oblap}
\Delta^{-1}u(t,x)=-\sum_{k\in\mathbb{Z}^n\backslash\{0\}}\frac{u_k(t)}{|k|_e^2}e^{i(k,x)}.
\end{equation}
Let $D$ be an open subset of $\Upsilon (T)$ and compact $K$ belongs to $D$.
Then formula (\ref{oblap}) involves an estimate:
$$
\|\Delta^{-1}u\|_{K}\le c_{K,D}\|u\|_{D},$$
where $c_{K,D}$ is a positive constant.

Define a set of linear operators
$S^t:\mathcal{O}(\Upsilon (T))\to\mathcal{O}(\Upsilon (T))$ by the formula:
$$S^tu(t,x)=\sum_{k\in\mathbb{Z}^n}u_k(t)e^{i(k,x)-\nu |k|^2_et},
$$
where  $\nu$ is the same constant that is in equation (\ref{N-S-1}).

Operators $\{S^t\}$ are bounded:
\begin{equation}
\label{yyy}
\begin{aligned}
\|S^tu\|_K&\le c_{K,D} \|u\|_D.
\end{aligned}
\end{equation}
Compact $K$ is in domain $D$, constant $c_{K,D}$ is positive.

Consider an initial problem for the equation:
\begin{equation}
\label{lap}
\begin{array}{rcl}
u_t&=&\nu\Delta u+f(t,x),\\
u\mid_{t=0}&=&\hat u(x)\in \mathcal{H}_{s}.
\end{array}
\end{equation}

The function $f$ belongs to the space $\mathcal{O}(\Upsilon (T))$
and $f(0,x)\in\mathcal{H}_{s}$, furthermore $\|f(t,x)-f(0,x)\|_s\to 0$
when $t$ vanishes.

It is easy to check that problem (\ref{lap}) has the unique solution
$u(t,x)\in\mathcal{O}(\Upsilon (T))$ that is
periodical in $x$ and it presents in a form:

\begin{equation}
\label{Lap_sol}
u(t,x)=S^t\hat u+\int\limits_{L(t)}S^{t-\tau}f(\tau,x)\,d\tau,
\end{equation}
where a contour $L(t)$ is constructed as follows:
$$
L(t)=\left\{s+i\frac{\Im t}{\Re t}s\mid 0\le s\le \Re t\right\}.
$$

\subsection{Majorant functions}\label{majjjj}
Let
$$
\begin{aligned}
v(t,x)&=&\sum_{k\in\mathbb{Z}^n}v_k(t)e^{i(k,x)}\in  \mathcal{O}(\Upsilon (T)),\\
V(\tau,x)&=&\sum_{k\in\mathbb{Z}^n}V_k(\tau)e^{i(k,x)}\in
C(I_T,\mathcal{H}_{s}),\\&&I_T=[0,T].\end{aligned}
$$
A notation $v\ll V$
means that  $|v_k(t)|\le V_k(\Re t)$ holds for all admissible $t$ and
$k\in \mathbb{Z}^n$.

If $u,U$ are vector-functions then a relation $u\ll U$ means that each component of
the vector $U$ majorates corresponding component of the vector $u$.

Define the following operators:
$$ Du=\sum_{k\in \mathbb{Z}^n}|k|u_ke^{i(k,x)},\quad
\Lambda^tu=\sum_{k\in \mathbb{Z}^n}u_ke^{i(k,x)-|k|_et\nu/2}.$$

Enumerate main properties of the operation "$\ll$".  Let
$u(t,x)\ll U(\tau,x)$ and $v(t,x)\ll V(\tau,x)$ then:
$$\begin{array}{rclrcl}
u+v&\ll&U+V,&uv&\ll&UV,\\
\lambda u&\ll& |\lambda|U,&\int\limits_{L(t)}u(s,x)\,ds&\ll&
2\int\limits_{0}^{\mathrm{Re}\, t}U(s,x)\,ds,\\
\partial_l u&\ll& DU,& S^tu&\ll& U,\\
D(uv)&\ll&UDV+VDU,& \Delta^{-1}u&\ll& U.
\end{array}$$
In these formulas we imply: $\lambda\in \mathbb{C}$.

Another property of "$\ll$" is as follows: there exists some positive
constant $c$ such that an estimate
\begin{equation}
\label{ewwrwr}
\Delta^{-1}\partial_j\partial_l u\ll c\,U
\end{equation}
holds for all functions $u\ll U$. Indeed,
expanding the left- and the
right-hand  side of (\ref{ewwrwr}) to Fourier series by formula (\ref{oblap})
we see
that the estimate follows from inequality:
$$|k_j k_l|\le c|k|_e^2,\quad k\in \mathbb{Z}^n.$$

Consider maps
$$f:\mathcal{O}(\Upsilon (T))\to \mathcal{O}(\Upsilon (T)),\quad
F:C(I_T,\mathcal{H}_s)\to C(I_T,\mathcal{H}_s).$$ We
say that the map $F$ majorates the map $f$ (denote by $f\ll F$) if for any functions $u,U$
the relation $u\ll U$ involves $f(u)\ll F(U)$.

\subsection{Existence Lemma}
Let $L_2(I_T,\mathcal{H}_s)$ be a space of maps $f$ from $I_T$ to
$\mathcal{H}_s$ that have square integrable norm: $\|f(t,\cdot)\|_s^2\in
L_2(I_T)$. The norm in $L_2(I_T,\mathcal{H}_s)$ is defined as usual:
$$(\|f\|^{L}_s)^2=\int_0^T\|f(t,\cdot)\|_s^2\,dt.$$
Let $u(x)=\sum_{k\in \mathbb{Z}^n}u_ke^{i(k,x)}\in \mathcal{H}_s$.
Define a semigroup $\{P_{\rho}^\lambda\}_{\lambda\ge 0}$ by the formula:
$$P_\rho^\lambda u=\sum_{k\in \mathbb{Z}^n}u_ke^{i(k,x)-\lambda |k|_e^2\rho},\quad \rho>0.$$
Define a map
$$\Phi(u,v)=\int_0^tP_\rho^{t-\xi}D(u(\xi,x)v(\xi,x))\,d\xi.$$
\begin{lem}
\label{first_lem}
The map $\Phi$ takes the space $L_2(I_T,\mathcal{H}_s)\times
L_2(I_T,\mathcal{H}_s)$ to the space $L_2(I_T,\mathcal{H}_s),$
and
\begin{equation}
\label{est_firs}
\|\Phi(u,v)\|^{L}_s\le c\|u\|^{L}_s\|v\|^{L}_s,
\end{equation}
a positive constant $c$ does not depend on $u,v$.
\end{lem}
{\it Proof.} The Lemma follows from the following
chain of estimates:
$$
\begin{aligned}
(\|\Phi(u,v)\|^{L}_s)^2&=\\
&=\int_0^T\Big\|\int_0^tP_\rho^{t-\xi}D(u(\xi,x)v(\xi,x))\,d\xi\Big\|_s^2\,dt\\
&\le
\int_0^T\,dt\int_0^t\,d\xi\int_0^t\,d\eta\|P_\rho^{t-\xi}D(uv(\xi,\cdot)\|_s
\|P_\rho^{t-\eta}D(uv(\eta,\cdot)\|_s\\
&= \Big(\int_0^T\,d\xi\int_0^\xi\,d\eta\int_\xi^T\,dt+
     \int_0^T\,d\eta\int_0^\eta\,d\xi\int_\eta^T\,dt\Big)\\
&\cdot\|P_\rho^{t-\xi}D(uv(\xi,\cdot)\|_s\|P_\rho^{t-\eta}D(uv(\eta,\cdot)\|_s\\
&\le c\Big(\int_0^T\|uv(\xi,\cdot)\|_s\,d\xi\Big)^2
\le c(\|u\|_s^{L})^2(\|v\|_s^{L})^2.
\end{aligned}
$$
The positive constant $c$ is taken as follows:
$(|k|_e^2+|j|_e^2)c\ge |k||j|.$

The last inequality is  the H\"{o}lder inequality.
To obtain the inequality before the last one, we termwise
integrate the series  which the norms are.
Admissibility of this chain of estimates follows from the standard
theorems of analysis.

Lemma is proved.

Recall that the
Sobolev space $W^1_2(I_T,\mathcal{H}_s)$ defines as a completion of the space
$C^\infty(I_T,\mathcal{H}_s)$ with respect to the norm
$$\|u\|_s^W=\max\{\|u\|^{L}_s,\|u_t\|^{L}_s\}.$$

Define by $\tilde W^1_2(I_T,\mathcal{H}_s)$ a subspace of
$W^1_2(I_T,\mathcal{H}_s)$ that consists of such a type functions
$u(0,x)=0$.

\begin{lem}
\label{second_lem}
The map $\Phi$ takes the space $\tilde W^1_2(I_T,\mathcal{H}_s)\times
\tilde W^1_2(I_T,\mathcal{H}_s)$ to the space $\tilde W^1_2(I_T,\mathcal{H}_s),$
and
\begin{equation}
\label{est_f}
\|\Phi(u,v)\|^{W}_s\le c\|u\|^{W}_s\|v\|^{W}_s,
\end{equation}
a positive constant $c$ does not depend on $u,v$.
\end{lem}

{\it Proof.} Note that
$$ \frac{dP_\rho^{t-\xi}}{dt}=-\frac{dP_\rho^{t-\xi}}{d\xi}.$$
Using this formula we derive:
$$
\begin{aligned}
\Phi_t(u,v)&=D(uv)-\int_0^t\frac{dP_\rho^{t-\xi}}{d\xi}D(uv(\xi,x))\,d\xi\\
&=D(uv)-P_\rho^{t-\xi}D(uv)\mid^t_0+\int_0^tP_\rho^{t-\xi}D(uv(\xi,x))_\xi\,d\xi\\
&=P_\rho^tD(uv)+\int_0^tP_\rho^{t-\xi}D(uv(\xi,x))_\xi\,d\xi.\end{aligned}
$$
Prepare  calculations:
$$\begin{aligned}
(\|\Phi_t(u,v)\|^L_s)^2&=\\
&=\int_0^T\Big\|P_\rho^tD(uv)(t,\cdot)+
\int^t_0P_\rho^{t-\xi}D(uv)_\xi(\xi,\cdot)\,d\xi\Big\|^2_s\,dt\\
&\le\int^T_0\|P_\rho^tD(uv)(t,\cdot)\|^2_s\,dt\\&+
2\int_0^T\|P_\rho^tD(uv)(t,\cdot)\|^2_s\,dt\int^T_0
\Big\|\int_0^tP_\rho^{t-\xi}D(uv)_\xi(\xi,\cdot)\,d\xi\Big\|^2_s\,dt\\
&+\int^T_0
\Big\|\int_0^tP_\rho^{t-\xi}D(uv)_\xi(\xi,\cdot)\,d\xi\Big\|^2_s\,dt.
\end{aligned}$$

The last term in this formula estimates by Lemma
\ref{first_lem}:
$$
\int^T_0
\Big\|\int_0^tP_\rho^{t-\xi}D(uv)_\xi(\xi,\cdot)\,d\xi\Big\|^2_s\,dt\le
c(\|u\|^W_s)^2(\|v\|^W_s)^2.$$
To estimate another term note that
$$uv(t,x)=\int_0^t(uv)_\xi\,d\xi,$$
thus we have
$$
\begin{aligned}
\int_0^T\|P_\rho^tD(uv)(t,\cdot)\|_s^2\,dt&=\int_0^T
\Big\|\int_0^tP_\rho^tD(uv)_\xi(\xi,\cdot)\,d\xi\Big\|_s^2\,dt\\
&\le
c(\|u\|^W_s)^2(\|v\|^W_s)^2.
\end{aligned}$$
Last estimate obtains in the same way that  Lemma \ref{first_lem} does.

Lemma is proved.

\begin{lem}
\label{exiseee}
For all positive constants $a,\rho$ and for every function $\hat V\in \mathcal{H}_s$
there is a positive constant $T$ such
that the equation
\begin{equation}
\label{ex_eqr}
V(t,x)=\hat V(x)+a\int_0^tP_\rho^{t-\xi}D(V^2)(\xi,x)\,d\xi,\quad \hat V\in
\mathcal{H}_s,
\end{equation}
 has an unique solution $V(t,x)\in C(I_T,\mathcal{H}_s)$.

For any sufficiently small $a$ the previous assertion remains valid for
$T=\infty$.
 \end{lem}

{\it Proof.} Since the space $W^1_2(I_T,\mathcal{H}_s)$
continuously embeds in $C(I_T,\mathcal{H}_s)$ (by the Sobolev embedding
theorem)
it sufficient to prove that equation (\ref{ex_eqr}) has a
solution in $W^1_2(I_T,\mathcal{H}_s)$.

After a change of variable: $V\to U=V-\hat V$ equation
(\ref{ex_eqr}) takes the form:
$$U=F(U)=a\int_0^tP_\rho^{t-\xi}D(U^2+2U\hat V+\hat V^2)(\xi,x)\,d\xi.$$
By Lemma \ref{second_lem} the operator $F$ is contracting in the
space $\tilde W^1_2(I_T,\mathcal{H}_s)$
for sufficiently small $T$.

If $a$ is sufficiently small the operator $F$ is also contracting.

Lemma is proved.

\section{Proof of theorem \protect\ref{main_theo}}
We shall prove Theorem \ref{main_theo} by the Majorant functions method. Namely,
system (\ref{N-S-1})-(\ref{N-S-11}) will be changed by another so called
majorant equation. Then we shall prove a theorem of existence for the
majorant equation and show that it involves the existence theorem for
original system.

The majorant functions method was originated by Cauchy and
Weierstrass and applied by Kovalevskaya
to prove an existence of analytic solutions in initial problems for PDE.
Further studies and applications of this technique contains in
\cite{lednev},\cite{Treshev3},\cite{zu}.
\subsection{The existence}
\label{perturbbb}
Now prove the first part of the theorem, the second one turns in the same
way with respect to the last assertion of Lemma \ref{exiseee}.

 Following standard procedure we
take operator $\mbox{div}$ from the right- and the left-hand sides of
equation (\ref{N-S-1}). Using equation (\ref{N-S-12}) we get
$\partial_i\partial_j (v^iv^j)=-\Delta p,$
where we summarize by the
repeated subscripts. So,
\begin{equation}
\label{pres}
p=-\Delta^{-1}\partial_i\partial_j (v^iv^j).\end{equation}
Substituting this formula to equation (\ref{N-S-1}) we obtain the
following problem:
\begin{equation}
\label{N-S_main}
\begin{aligned}
(v^k)_t&=A^k_l\partial_j(v^jv^l)+\nu\Delta
v^k,\quad A^k_l=(\Delta^{-1}\partial_k\partial_l-\delta_{kl}),\\
v^k\mid_{t=0}&=\hat{v}^k,
\end{aligned}
\end{equation}
where $\delta_{kl}=1$ for $k=l$ and $0$ otherwise.

Operator $A^k_l$ is bounded:
\begin{equation}
\label{yyz}
\begin{aligned}
\|A^k_lu\|_K&\le c_{K,D} \|u\|_D,
\end{aligned}
\end{equation}
Compact $K$ belongs to the domain $D$, constant $c_{K,D}$ is positive.

According to formula (\ref{Lap_sol}) we present equation (\ref{N-S_main}) in
the form:
\begin{equation}
\label{plmnjk}
\begin{aligned}
(v^k)(t,x)&=G^k(\mathbf{v})=S^{t}\hat
v^k(x)+\int_{L(t)}S^{t-\xi}A^k_l\partial_j(v^jv^l)(\xi,x)\,d\xi,\\
G(\mathbf{v})&=(G^1,\ldots, G^n).
\end{aligned}
\end{equation}

\begin{lem}Let $\hat v^k\ll \hat V$. There exist positive constants
$\rho,a$
such that
if $V(t,x)$ is the solution of equation (\ref{ex_eqr}) corresponding to these constants
 then
the map $G$ takes a set
$$W=\{u\in \mathcal{O}(\Upsilon (T))\mid u\ll
\Lambda^{\mathrm{Re}\, t}V,\quad\mathrm{div}\, u=0\}$$
into itself.
\end{lem}

{\it Proof.}
It easy to check that the map $G$ takes a solenoidal vector-field to
a solenoidal vector-field.

Choose the constant $a$ such that
if $$v^k\ll \Lambda^{\mathrm{Re}\,t}V$$ then
 $$G^k(\mathbf{v})\ll\Lambda^{\mathrm{Re}\,t}
  \hat V+a\int_0^{\mathrm{Re}\,t}S^{\mathrm{Re}\,t-\xi}
 D(\Lambda^{\xi}V)^2\,d\xi.$$
 Choose the constant $\rho$ such that the estimate holds:
 $$
\Lambda^{\tau}
  \hat V+a\int_0^{\tau}S^{\tau-\xi}
 D(\Lambda^{\xi}V)^2\,d\xi\ll \Lambda^{\tau}
 \Big(\hat V+a\int_0^{\tau}P_\rho^{\tau-\xi}
 D(V)^2\,d\xi\Big)=\Lambda^\tau V.$$

Lemma is proved.

\begin{lem}
\label{gggggk}The set $W$ is convex and it is compact in $\mathcal{O}(\Upsilon (T))$.
\end{lem}
{\it Proof.} The convexity is obvious.

According to the Montel theorem it sufficient to prove that the set $W$ is
bounded. For $(t,x)\in\Upsilon_r(T)$ we have
$$
\begin{aligned}
|v^k(t,x)|&\le\Big|\sum_{j\in\mathbb{Z}^n}v^k_j(t)e^{i(j,x)}\Big|\le
\sum_{j\in\mathbb{Z}^n}|v^k_j(t)|e^{|j|r/\alpha}\\&\le
\sum_{j\in\mathbb{Z}^n}V_j(\Re t)e^{-\frac{\nu}{2}|j|_e\mathrm{Re}\,t+|j|r/\alpha}
\le\sum_{j\in\mathbb{Z}^n}V_j(\Re t)e^{-\frac{\nu}{2}|j|_er+|j|r/\alpha},
\end{aligned}
$$
where
$$\alpha=\frac{3}{\nu\underline{c}}.$$
Thus \begin{equation}
\label{po}\|v^k\|_{\Upsilon_r(T)}\le
\sup_{t\in I_T}\sum_{j\in\mathbb{Z}^n}
V_j(\Re t)e^{-\frac{\nu}{2}|j|_er+|j|r/\alpha}.\end{equation}
The right-hand side of  estimate (\ref{po}) is bounded for all admissible
$r$.
It follows from the fact that $V\in
C(I_T,\mathcal{H}_s).$

Lemma is proved.

The map $G$ is continues with respect to the seminormed topology in
$\mathcal{O}(\Upsilon (T))$. Thus according to the Theorem \ref{main_t} and
Lemma \ref{gggggk}
 it has a fixed point $\mathbf{v}(t,x)\in W$.  This fixed point is
 solution of equations (\ref{N-S-1}), (\ref{N-S-12}) for $(t,x)\in \Upsilon (T)$.

Let us show that the vector-field $\mathbf{v}(t,x)$ satisfies the
conditions (\ref{poiuyt}). (We check these equalities just for $k=0$.
The others  turns out in analogous way.)

So,
$$
\|v^k(t,\cdot)-\hat v^k\|_s\le
\Big\|\int_{L(t)}S^{t-\xi}A^k_l\partial_j(v^jv^l)(\xi,x)\,d\xi\Big\|_s+
\|\hat v^k-S^{t}\hat
v^k\|_s.$$

The second term in this formula obviously vanishes when $t\to 0$. The
first one estimates in the following way:
$$
\begin{aligned}
\Big\|\int_{L(t)}S^{t-\xi}A^k_l\partial_j(v^jv^l)(\xi,x)\,d\xi\Big\|_s
\le a\Big\|\int_0^{\mathrm{Re}\,t}P_\rho^{\mathrm{Re}\,t-\xi}D(V)^2(\xi,x)\,d\xi\Big\|_s\to
0,\\ t\to 0.
\end{aligned}
$$

The existence is proved.

\subsection{The Uniqueness}
Let $\mathbb{T}^n_r=\{x\in \mathbb{C}^n\mid\Re x\in\mathbb{T}^n,\quad |\Im
x|_m<r\}$ be a complex neighborhood of the torus.
Let $u=\sum_{k\in \mathbb{Z}^n}u_ke^{i(k,x)}\in \mathcal{O}(\mathbb{T}^n_r)$,
consider a norm
$\|u\|^*_{r'}=\sum_{k\in \mathbb{Z}^n}|u_k|e^{|k|r'},\quad r'<r.$

This norm satisfies to the Cauchy inequality: $$\|u\|^*_r\le
\frac{c}{\delta}\|u\|^*_{r+\delta}.$$

In this section we consider only real value of variable $t$.

We prove that problem (\ref{N-S_main}) has an unique solution. Assume the
converse. Let $v^j(t,x)$ and $u^j(t,x)$ be a different solutions of this
problem for $t$ greater than some positive constant $\hat t$:
$$K=\sup_{\hat t\le t\le T}\|v^j(t,x)-u^j(t,x)\|^*_{\tilde r}>0,\quad \alpha\tilde r< \hat t.$$

\begin{lem}
\label{poipppl}
There is a constant $C$ such that for all positive integers $k$ we have
\begin{equation}
\label{kjhggf}
\|v^j(t,x)-u^j(t,x)\|^*_{r}\le K\Big(\frac{Ce(t-\hat t)}{\tilde
r-r}\Big)^k,\quad r<\tilde r.
\end{equation}
\end{lem}

Tending $k\to\infty$ in (\ref{kjhggf}) we obtain
$$\|u^j-v^j\|^*_r=0,\quad\mathrm{for}\quad \hat t\le t< \frac{\tilde
r-r}{Ce}+\hat t.$$ This contradiction proves the uniqueness.

{\it Proof of Lemma \ref{poipppl}}
By formula (\ref{plmnjk}) we get
$$
u^m-v^m=\int_{\hat t}^tS^{t-s}A^m_l\partial_j(v^jv^l-u^ju^l)\,ds.
$$
So for $k=1$ formula (\ref{kjhggf}) follows from the Cauchy inequality and formulas
(\ref{yyy}),
(\ref{yyz}):
$$
\|u^j-v^j\|_r^*\le \frac{C}{\tilde r-r}\|u^j-v^j\|_{\tilde r}^*$$
Suppose it is valid for
$k$ and prove it for $k+1$.

For $r<r'<\tilde r$ we have
$$\begin{aligned}
\|u^j-v^j\|^*_r&\le \int\limits^t_{\hat t}\frac{C}{r'-r}\|u^j-v^j\|^*_{r'}\,d\tau\le
\frac{CK}{r'-r}\int\limits^t_{\hat t}\left(\frac{Ce\tau}{\tilde{r}-r'}\right)^k\,d\tau
\\&\le\left.\frac{CK(Me)^k(t-\hat t)^{k+1}}{(r'-r)(\tilde{r}-r')^k(k+1)}
\right|_{r'=r+\frac{\tilde{r}-r}{k+1}}\le\left(\frac{Ce(t-\hat t)}{\tilde{r}-r}\right)^{k+1}K.
\end{aligned}$$

Lemma is proved.

\section{Generalized Schauder's fixed point theorem}\label{schaussect}

Let $(L,\{\|\cdot\|_\omega\}_{\omega\in\Omega})$ be a seminormed space.
A basis of topology in $L$ is given by the balls:
$$B_\tau(\hat{x},r)=\{x\in L\mid \|x-\hat{x}\|_\tau< r\}.$$
And let there exists
$\omega'\in \Omega$ such that $\|\cdot\|_{\omega'}$ is a norm.
Compact set $K\subset L$ is convex.

Consider a continuous map $f:K\to K$.

\begin{theo}[Generalized Schauder's theorem]
\label{main_t}
There exists a point $\hat x\in K$ such that $f(\hat x)=\hat x$.
\end{theo}

This result is well known. It is a special case of theorem from \cite{Browder}.
Nevertheless for completeness of exposition we present direct and simple proof.

Recall the original formulation of Schauder's theorem \cite{Nirenberg}.
Let $(L,\|\cdot\|)$ be a Banach space and $K\subset L$
 be a convex compact set.
Then a continuous map $f:K\to K$ has a fixed point $\hat x\in K$.

Note that though this formulation
include completeness of the space, actually
this condition is  not necessarily. The point is that
the prove of this theorem (see \cite{Nirenberg}) considers the map $f$
only on the compact $K$ but any compact set is complete and can be
embedded to a completion of the space $L$.

\subsection{Proof of Theorem \protect{\ref{main_t}}}

Let $(E,\{\rho_\omega\}_{\omega\in\Omega})$ and
$(F,\{d_\sigma\}_{\sigma\in\Sigma})$ are semimetric spaces and there exist
$\omega',\sigma'$ such that the
semimetricsis $\rho_{\omega'}$ and
$d_{\sigma'}$ are metricsis.

Consider a compact (with respect to the semimetric topology)
  $K\subset (E,\{\rho_\omega\}_{\omega\in\Omega})$
 and a map $f:E\to F$.
\begin{lem}
\label{equ}
If the map $f:E\to F$ is continuous on $K$
with respect to the semimetric topology then it
is continuous on $K$ as a map of the metric space $(E,\rho_{\omega'})$
 to the
metric space $(F,d_{\sigma'})$.
\end{lem}

{\it Proof.}
Let $\{x_n\}\subset K$ be a sequence such that $\rho_{\omega'}(x_n,a)\to
0$ as $n\to \infty$ where $a\in K$ and we put $y_n=f(x_n)$.
So we must prove that
$d_{\sigma'}(y_n,b)\to 0$ where $b=f(a)$.

Assume the converse.
Then there exists a subsequence $\{y'_n\}\subseteq \{y_n\}$ such that
$d_{\sigma'}(y'_n,b)\ge c>0$. A set $\hat K=f(K)$ is  compact as
an image of
a compact set under a continuous map and $\{y'_n\}\subset \hat K$.
Thus, there
 exists a subsequence
$\{y''_n\}\subseteq\{y'_n\}$
such that
\begin{equation}
\label{sdf}
d_{\sigma}(y''_n,\beta)\to 0,\quad\sigma\in \Sigma,\quad
\beta\ne b.\end{equation}

Let $\{x''_n\}\subseteq \{x_n\}$ be a sequence such that
$y''_n=f(x''_n)$. Consider a subsequence $\{x'''_n\}\subseteq \{x''_n\}$
that converges with respect to the semimetric topology:
$\rho_\omega(x'''_n,a)\to 0$ for all $\omega\in \Omega$ and let
$y'''_n=f(x'''_n)$. Note that $\{y'''_n\}\subseteq\{y''_n\}$.

Since $f$ is continuous we have $d_\sigma(y'''_n,b)\to 0$ for all
$\sigma\in\Sigma$. On the other hand we have
(\ref{sdf}).
This contradiction proves the Lemma.

Lemma is proved.

Theorem \ref{main_t} almost directly follows from
original Schauder's theorem
and Lemma \ref{equ}. Indeed, by Lemma \ref{equ} the map $f$ is continuous
 on $K$ with respect to the norm $\|\cdot\|_{\omega'}$.
By $\overline{L}$ denote a
completion of $L$ with respect to the same norm.

It is easy to check that the compactness of the set $K$ with
respect to the seminormed topology involves the compactness of $K$ with
respect to the norm $\|\cdot\|_{\omega'}$.
So we obtain the continuous map $f:K\to K$ where $K$ is a convex compact
set
in the Banach space $\overline{L}$.

By  original Schauder's theorem we get the fixed point $\hat x$.

Theorem \ref{main_t} is proved.

 \end{document}